\documentclass{appolb}

\usepackage{amsfonts,amsmath,amssymb,bm}
\usepackage{graphicx}
\usepackage{dcolumn}

\begin{document}
\title{Complexity characteristics of currency networks}
\author{A. Z. G\'orski$^1$, S. Dro\.zd\.z$^{1,2}$, J. Kwapie\'n$^1$ and
P. O\'swi\c ecimka$^1$
\address{$^1$Polish Academy of Sciences, Institute of Nuclear Physics,
Radzikowskiego~152, Krak\'ow PL 31-342, Poland, \\%
$^2$University of Rzesz\'ow, Institute of Physics,%
Rzesz\'ow PL 35-310, Poland}}
\maketitle
\begin{abstract}
\parbox{14cm}

A large set of daily FOREX time series is analyzed. The corresponding
correlation matrices (CM) are constructed for USD, EUR and PLZ used as the
base currencies. The triangle rule is interpreted as constraints reducing
the number of independent returns.
The CM spectrum is computed and compared with the cases of shuffled currencies
and a fictitious random currency taken as a base currency.
The Minimal Spanning Tree (MST) graphs are calculated and the clustering
effects for strong currencies are found. It is shown that for MSTs the node
rank
has power like, scale free behavior. Finally, the scaling exponents are
evaluated and found in the range analogous to those identified recently for
various complex networks. 
\end{abstract}

%
%
%
\newpage

\section{Introduction}

 Analysis of correlations among financial assets is of great interest for
practical, as well as for fundamental reasons. Practical aspects are mainly
related to the theory of optimal portfolios \cite{PortfolioManagement}.
The theoretical interest results from the fact that such study may shed more
light on the universal aspects of complex systems organization.
The world currency network can definitely be considered as complex. 

 In this paper we analyze daily FOREX (FX) time series of 60 currencies
(including gold, silver and platinum) from the period Dec 1998 -- May 2005,
provided by University of British Columbia \cite{datasource}. The $5\sigma$
filter was applied to avoid spikes due to errors.

For a value $x_i(t)$ of the $i$th asset ($i = 1, \ldots, N$) at time $t$, one
defines its return $G_i(t)$ as
\begin{equation}
\label{DEFreturn}
G_i(t) = \ln x_i(t+\tau) - \ln x_i(t) \simeq
\frac{ x_i(t+\tau) - x_i(t)} {x_i(t)} \ ,
\end{equation}
where the return time $\tau$ is also called the time lag. The normalized
returns, $g_i(t)$ are defined as
\begin{equation}
\label{DEFnormret}
g_i(t) = [ G_i(t) - \langle G_i(t) \rangle_T ] / \sigma(G_i) \ ,
\end{equation}
where $\langle\ldots\rangle_T$ denotes averaging over variable $t$ with
the averaging window $T$ and
$\sigma(G_i)$ is the standard deviation (volatility) of $G_i$.

 The stock market time series $x_i(t)$ are always expressed in terms of the
local currency. However, for the FX data we have exchange rates, instead.
Denoting currencies by $n$ consecutive capital Latin letters $A, B, C, \ldots$
the corresponding FX data $x_i(t)$ can be expressed as their quotients:
$x_{AB}(t) = A(t)/B(t)$.
Neglecting friction caused by fees (this is usually negligible in open market
transactions) one obtains two types of constraints among $n$ currencies
\begin{equation}
\label{TriangleEffect}
\frac{A(t)}{B(t)} \; \frac{B(t)}{A(t)} = 1 \ , \ \ \
\frac{A(t)}{B(t)} \; \frac{B(t)}{C(t)} \; \frac{C(t)}{A(t)} = 1 \ ,
\end{equation}
where the second constraint is called the triangle rule \cite{McDonald2005}.
Eqs. (\ref{TriangleEffect}) can be rewritten in terms
of returns that gives the following identities
\begin{equation}
\label{TriangleEffectG}
\begin{split}
& G_{AB}(t) = - G_{BA}(t) \\
& G_{AB}(t) + G_{BC}(t) + G_{CA}(t) = 0 \ .
\end{split}
\end{equation}

For $n$ currencies there are in principle $n(n-1)$ possible exchange rates
$x(t)$ and corresponding returns $G(t)$.
Due to the first of eqs. (\ref{TriangleEffect}) half of them are simply related
to the
remaining values. The triangle effect can be shown to give additional
$(n-1)(n-2)/2$ independent constraints. This leaves us with $(n-1)$
independent exchange rates and returns for $n$ currencies,
$i = 1, \ldots, n-1$.
One currency can be chosen as a reference currency (denominators)
and we shell call it the {\it base currency}.
Taking different currencies as the base currency one can obtain a different
"picture" of the market though in principle all these pictures should
contain the same information.

 In this paper we construct correlation matrices (CMs) for the FX time series
and the corresponding Minimal Spanning Trees (MSTs). Finally, the scale free
distribution of node multiplicity is found and the corresponding scaling
exponents are estimated.
The complex network approach seems to be one of the most promising do
deal with such extremely complicated systems, as was suggested recently
\cite{AlbertBarabasi2002,Boccaletti2006}.

\section{Correlation matrices}

The correlation matrix (CM) $C_{ij}$ is defined in terms of returns
(\ref{DEFreturn})
as
\begin{equation}
\label{DEFcorrM}
C_{ij} =
\frac{ \langle G_i(t) G_j(t) \rangle_T - \langle G_i(t) \rangle_T \langle G_j(t)
\rangle_T } { \sigma(G_i) \sigma(G_j) } \ .
\end{equation}
The (symmetric) correlation matrix can also be computed in terms of the
normalized returns. To this end one has to form $N$ time series
$\{ g_i(t_0), g_i(t_0+\tau), \ldots, g_i(t_0+(T-1)\tau) \}$ of length
$T$. Hence, we can built an $N \times T$ rectangular matrix {\bf M}.
The correlation matrix (\ref{DEFcorrM}) can be written in matrix notation as
\begin{equation}
\label{corrMmatrix}
{\mathbf C} \equiv [ C ]_{ij} = \frac{1}{T} {\mathbf M} {\widetilde{\mathbf M}}
\ ,
\end{equation}
where tilde, $\widetilde{\mathbf M}$ stands for the matrix transposition.
To avoid artificial reduction of the rank of this matrix, one should have
sufficiently large time window for averaging: $T \ge N$.

\begin{figure}
\begin{center}
  \includegraphics[width=11.0cm,angle=0]{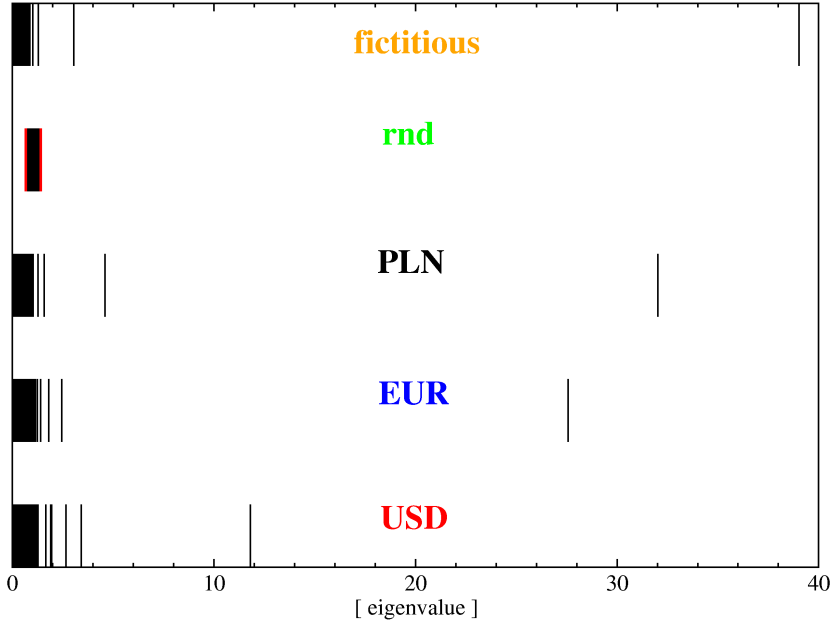}
  \caption{Eigenspectra of correlation matrices for USD, EUR, PLN, shuffled
USD and a random fictitious currency taken as the base currency, respectively.}
\end{center}
\label{fig:fig1}
\end{figure}

By construction the trace of a correlation matrix equals to the number of time
series
\begin{equation}
\label{CMtrace}
\text{Tr} \ {\mathbf C} = N \ .
\end{equation}
When some of the time series become strongly dependent, zero
eigenvalues emerge (zero modes).

The eigenspectrum of CM for USD, EUR and PLN as the base currency is plotted
in Fig~1. 
For comparison, two additional sets of time series were generated. 
As the first one, the USD based time series were taken and all of them were
randomly and independently shuffled. This set is denoted as (rnd). 
As all time correlations are destroyed the case (rnd) is clearly different than
all other cases. In particular, it is very close to the
random matrix spectrum, where the theoretical upper and lower limit for the
spectrum is given by \cite{SenguptaMitra}
\begin{equation}
\label{WishartLimit}
\lambda_{min} = 1 + \frac{1}{q} - \frac{2}{\sqrt{q}} \ , \quad
\lambda_{max} = 1 + \frac{1}{q} + \frac{2}{\sqrt{q}} \ ,
\end{equation}
where $q = T/N$. In our case eq.~(\ref{WishartLimit}) gives $\lambda_{min} =
0.67$ and $\lambda_{max} = 1.41$, in perfect agreement with the plot.

 In the second case, a fictitious currency was genereted with returns identical
to Gaussian uncorrelated noise and it was used as the base currency for our time
series. 
In this case time correlations of all other real currencies were preserved and
it is
denoted as "fictitious" (fict). The CM spectrum here is qualitatively similar to
real currencies. 

 For the real currencies the maximal eigenvalue is smallest for USD, larger
for EUR, much larger for PLZ and the largest for a fictitious random currency
taken as the base currency, respectively.
The magnitude of separation of the largest eigenvalue from $\lambda_{max}$
can be considered a measure of collectivity of the underlying dynamics
\cite{DrozdzWojcik}. 
Similar effects are observed for the stock market
correlations \cite{AZGStaszek2000}.

\section{Minimal Spanning Tree graphs}

Looking at large numerical matrices is not very enlightening. Instead, there are
useful visualizations that can be used for their analysis. In particular, the
Minimal Spanning Trees that were introduced in graph theory long ago
\cite{MSTBoruvka,MSTkruskal} and later rediscovered several
times \cite{MSTrefA,MSTrefB}. Recently they were 
applied to analyze the stock correlations \cite{MantegnaTree}.
Here, to draw the MST graph the following metric has been proposed
\begin{equation}
\label{MSTmetric}
d(i,j) = \sqrt{ 2 (1 - C_{ij}) } \ .
\end{equation}
Nodes corresponding to assets with the closest correlation coefficients are
successively linked with a line. As a result one obtains a tree-like
connected graph.

\begin{figure}
\begin{center}
  \includegraphics[width=12.0cm,angle=0]{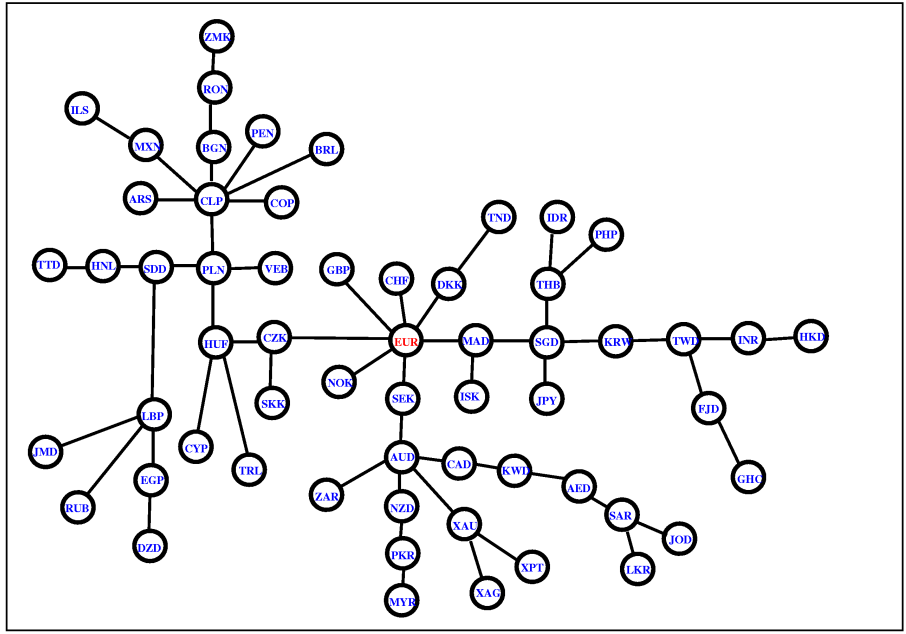}
  \caption{Minimal Spanning Tree for USD as the base currency. In absence of
USD only moderate size clusters are visible.}
\end{center}
\label{fig:fig2}
\end{figure}
 The corresponding MST graphs for USD, EUR and PLN  are shown in Figs.~2--4,
respectively. In Fig.~2 USD is absent and one can see nodes with relatively
small degree (small number of links). On the other hand, for EUR taken as the
base currency (Fig.~3) we have two large clusters --- USD and SAR cluster, 
both with high degree. The SAR cluster is present because of the strong
coupling of both currencies, USD and SAR. The last currency is artificially
fixed to USD. 
In Fig.~4 PLN is taken as the base currency. Here, we have a larger USD cluster
and smaller clusters, including the EUR cluster. The picture here is in a sense
intermediate. 
\begin{figure}
\begin{center}
  \includegraphics[width=12.0cm,angle=0]{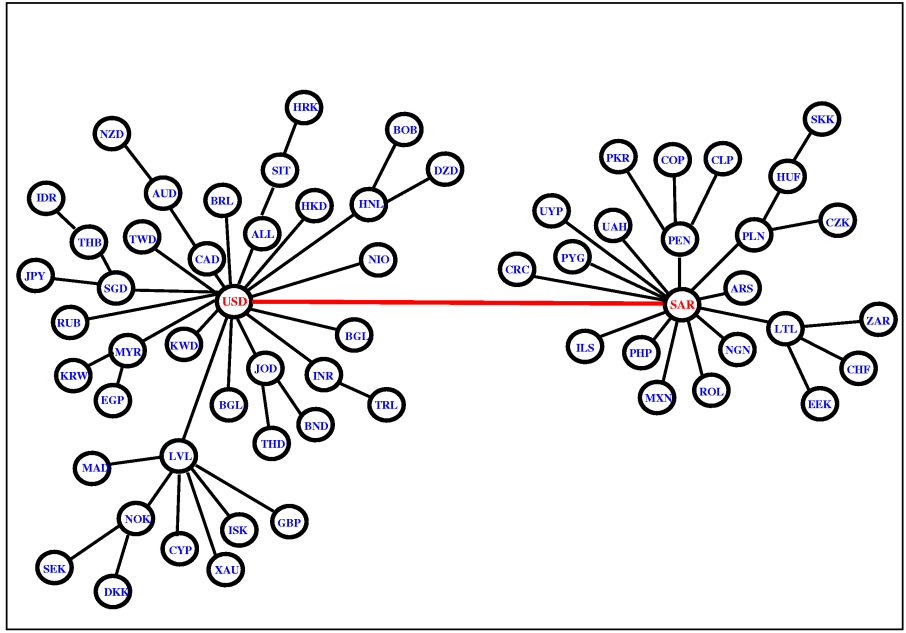}
  \caption{Minimal Spanning Tree for EUR as the base currency. USD and SAR are
in central positions of two large clusters.}
\end{center}
\label{fig:fig3}
\end{figure}
\begin{figure}
\begin{center}
  \includegraphics[width=12.0cm,angle=0]{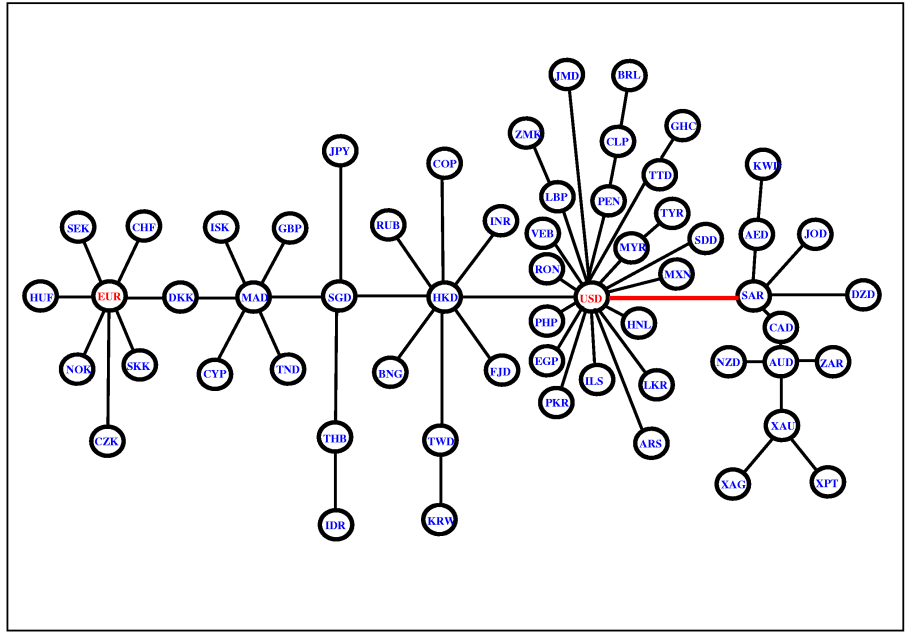}
  \caption{Minimal Spanning Tree for PLN as the base currency with large USD
cluster. Modest EUR cluster in the left part of the graph.}
\end{center}
\label{fig:fig4}
\end{figure}
 
We have also plotted MST for the correlation matrix with the USD as the base
currency, but all the corresponding currency return time series are shuffled
independently (Fig.~5). 
In this case all time correlations are killed. This corresponds to the (rnd)
spectrum in Fig.~1. In this case larger clusters are absent, as one can expect. 
Finally, for a fictitious (fict) randomly generated currency (a prototype of a
currency whose dynamics is completely disconnected from the rest) as the base
currency one obtains MST graph as in Fig.~6. Here, its structure is
qualitatively similar as for PLN taken as the base currency. 
This similarity even on a more quantitative level can be seen from Fig.~1.
\begin{figure}
\begin{center}
  \includegraphics[width=12.0cm,angle=0]{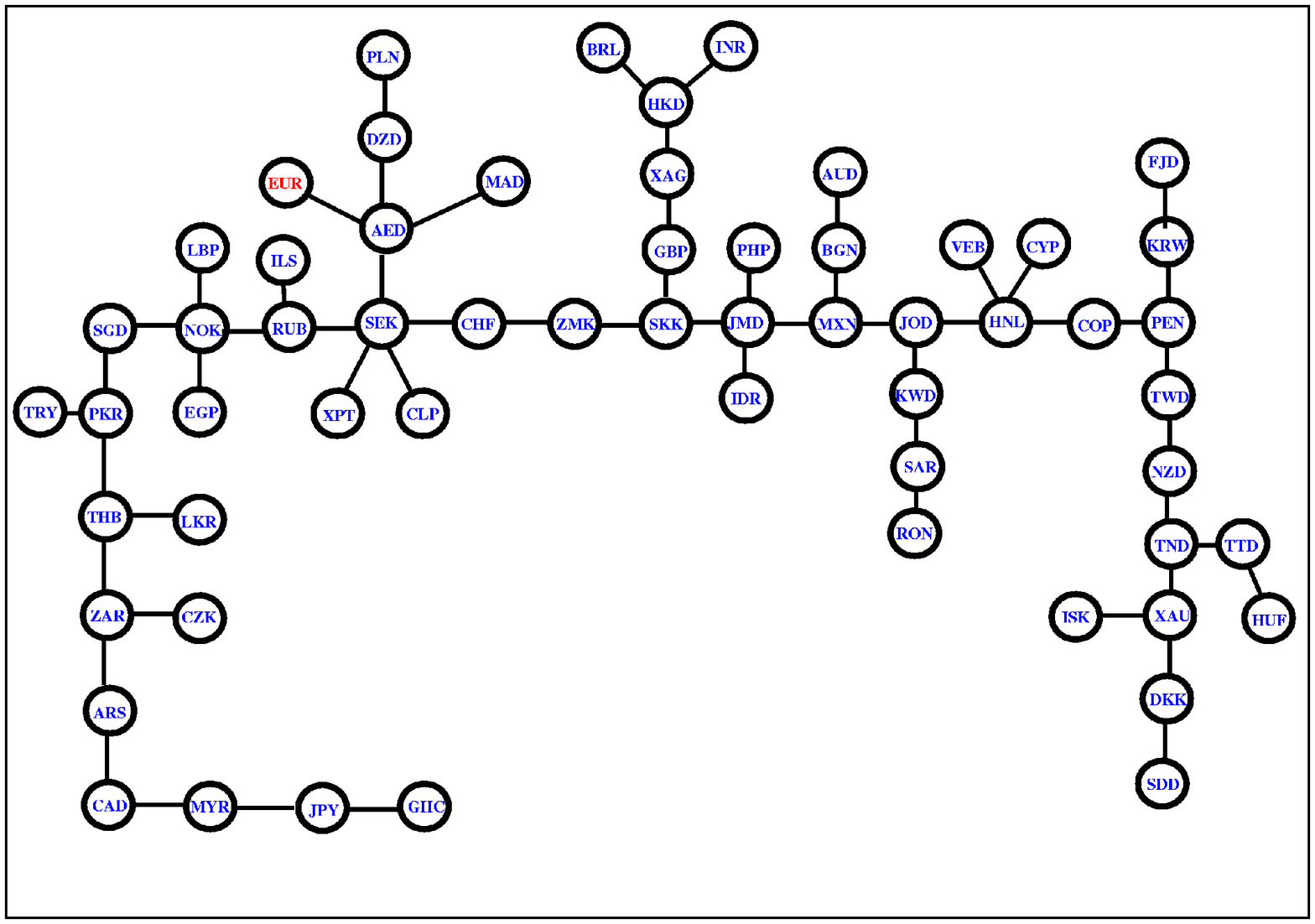}
  \caption{Minimal Spanning Tree for shuffled time series.}
\end{center}
\label{fig:fig5}
\end{figure}
\begin{figure}
\begin{center}
  \includegraphics[width=12.0cm,angle=0]{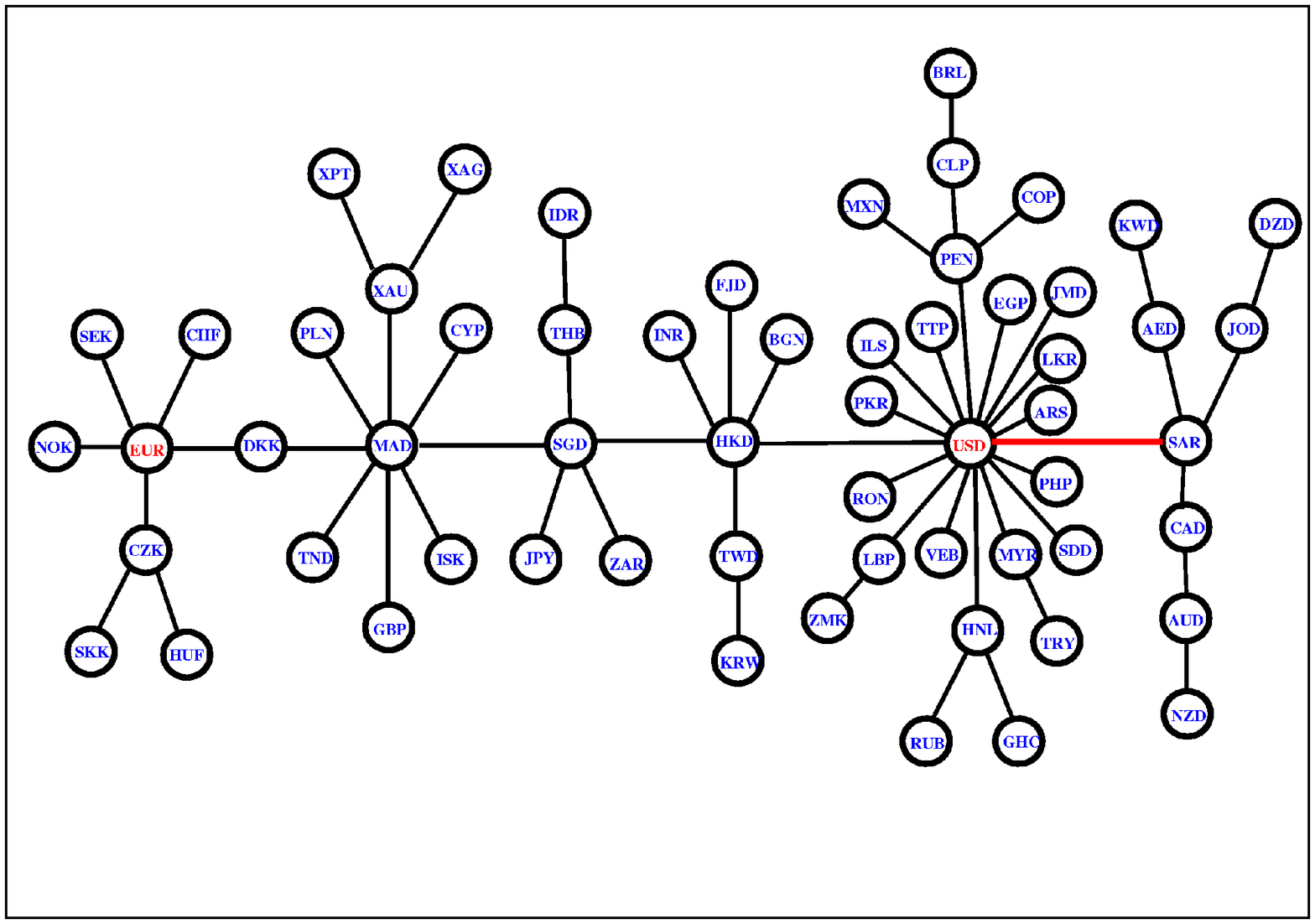}
  \caption{Minimal Spanning Tree for a fictitious Gaussian currency as
the base currency.}
\end{center}
\label{fig:fig6}
\end{figure}

\section{Power like scaling and conclusions}

 Because we have used considerably large number of currencies it is possible to
estimate the integrated distribution of the nodes' degree for all plots. 
The most interesting question is the type of this distribution. 
For complex networks it has been found that these distributions usually have 
scale free power like scaling. 
Indeed, we have found good power like scaling in all cases except for the
shuffled case, where all time correlations are wiped out. The log--log plot of
the integrated probability distribution for the nodes' degree (multiplicity) is
plotted in Fig.~7. The corresponding dashed lines represent the power like
fits. The numerical data are listed in Table~1. In addition to the scaling
exponent, $\alpha$, its standard error, relative error and Pearsons-r
coefficient are given. Except the shuffled case, standard error is of order of
a few percent and the r-coefficient is $>0.97$. This suggests a good
power like scaling. The largest error is for USD. In this case, power like fit
seems to be not so good. For the shuffled case, where time correlations are 
wiped out, one cannot see a power like scaling at all. 
The case of currencies expressed in terms of the USD 
seems to interpolate between the scale free and the shuffled cases. 
This may reflect the strong independence of the US currency. 
\begin{figure}
\begin{center}
  \includegraphics[width=11.0cm,angle=0]{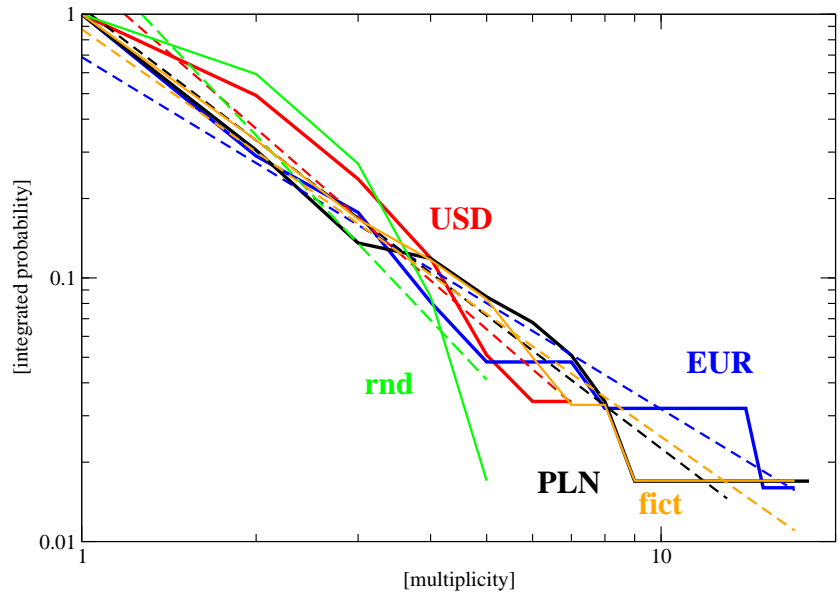}
  \caption{Integrated probability distribution of nodes' multiplicity for the
Minimal Spanning Tree graphs (Figs.2-6). The linear fits are represented by
corresponding dashed
lines.}
\end{center}
\label{fig:fig7}
\end{figure}

 Numerical results for all fits can be found in Table~1. It is worth to notice,
that, except the shuffled case, for all cases we have obtained
the scaling exponent in the range $1 < \alpha < 2$ (with average close to 1.6),
the same range as for the finite average
L\'evy stable distributions \cite{LevyDistr}. What is more important, with
rare exceptions, these exponents are similar to those found in different
complex networks, such as WWW pages ($\alpha=1.4$), physical internet networks
with nodes representing hosts (1.38), routers (1.18) and peer--to--peer networks
(1.19), protein--protein interaction network in the yeast (1.4), metabolic
reactins network (1.15), movie actor collaboration network (1.3), phone calls
(1.1), words co-occurence (1.7)
--- for references see \cite{AlbertBarabasi2002,Boccaletti2006}.

\begin{table}
\begin{center}
  \caption{\label{tab:table1} Numerical results for Minimal Spanning Trees
represented by Figs.2-6. $\alpha$, its standard and relative error and
Pearson's $r$ are given.}
      \begin{tabular}{|lcccc|}%
      \hline %
      \hline %
      \ base currency & $\alpha$ & std. error & \% & r--coeff. \\
      \hline%
      \hline%
        USD           & 1.913 & $\pm$0.183 & $\pm$9.6\% & 0.998 \\
     \ \ \ EUR        & 1.335 & $\pm$0.086 & $\pm$6.4\% & 0.970 \\
     \ \ \ PLN        & 1.488 & $\pm$0.084 & $\pm$5.7\% & 0.975 \\
     \ \ \ rnd        & 2.327 & $\pm$0.627 & $\pm$27 \% & 0.906 \\
     \ \ \ fictitious & 1.546 & $\pm$0.083 & $\pm$5.4\% & 0.979 \\%
     \hline
     \hline
\end{tabular}
\end{center}
\end{table}


\end{document}